\documentclass[a4paper,11pt]{article}
\usepackage[utf8]{inputenc}
\usepackage[margin=1.5cm]{geometry}
\usepackage{graphicx}
\usepackage{epsfig}
\usepackage{amsmath}
\usepackage{amssymb}
\usepackage{cite}
\usepackage{float}
\usepackage[font={footnotesize,it}]{caption}

\begin{document}

\begin{titlepage}
\title{Thermodynamic Geometry and Phase Transitions of Dyonic Charged AdS Black Holes}
\author{}
\date{
Pankaj Chaturvedi, Anirban Das$^{\dagger}$ and Gautam Sengupta
\thanks{E-mail:~ cpankaj@iitk.ac.in, anirbandas21989@gmail.com, sengupta @iitk.ac.in$^{*}$}
\vskip0.4cm
{\sl ${}^*$Department of Physics, \\
Indian Institute of Technology,\\
Kanpur 208016, India}
\vskip0.4cm
{\sl $^{\dagger}$Department of Theoretical Physics,\\
 Tata Institute of Fundamental Research,\\
Homi Bhabha Rd, Mumbai 400005, India}}
\maketitle
\abstract{
\noindent
We investigate phase transitions and critical phenomena of four dimensional dyonic charged AdS black holes in the framework of thermodynamic geometry. In a mixed canonical grand canonical ensemble with a fixed electric charge and varying magnetic charge these black holes exhibit liquid gas like first order phase transition culminating in a second order critical point similar to the Van der Waals gas. We show that the thermodynamic scalar curvature R for these black holes follow our proposed geometrical characterization of the R-crossing Method for the first order liquid gas like phase transition and exhibits a divergence at the second order critical point. The pattern of R crossing and divergence exactly corresponds to those of a Van der Waals gas described by us in an earlier work.}

\end{titlepage}

\section{Introduction}

In the last two decades the investigation of the thermodynamics of black holes has evolved into one of the most crucial issues in the study of quantum theories of gravity\cite{Hawking:1974sw,Bekenstein:1973ur, Bardeen:1973gs, Wald:1999vt,Page:2004xp, Ross:2005sc, Townsend:1997ku, Dabholkar:2012zz}\footnote {This is an extensive field, for detailed reviews see the references mentioned}. The thermodynamics of asymptotically Anti de Sitter (AdS) black holes has assumed critical importance in this context owing to the famed AdS-CFT duality.  Although a complete understanding of the microscopic statistical basis of black hole thermodynamics has eluded a clear explanation the study of the thermodynamics of AdS black holes has revealed a rich variety of phase structures and critical phenomena \cite {Hawking:1982dh,Chamblin:1999tk, Chamblin:1999hg,Caldarelli:1999xj}. Remarkably in \cite{Chamblin:1999tk,Chamblin:1999hg,Caldarelli:1999xj,Kubiznak:2012wp} an interesting similarity between the phase diagrams of electrically charged Reisner Nordstrom (RN) AdS  black holes and that of a Van der Waals fluid could be established. 

In a completely different context a geometrical approach to thermodynamics and phase transitions has been developed through the work of Weinhold \cite{Weinhold:1975sc, Weinhold:1975hp} and Ruppeiner \cite{Ruppeiner:1995zz}. In this framework a Riemannian geometry of classical thermodynamic fluctuations, with a Euclidean signature, may be defined for the equilibrium thermodynamic state space of any thermodynamic system.  The probability distribution of such fluctuations in a Gaussian approximation could then be related to the positive definite invariant interval defined by this geometry. It was found that the thermodynamic scalar curvature arising from this equilibrium state space  geometry encodes the microscopic interactions underlying the thermodynamic system. From standard scaling and hyperscaling arguments it was possible to show that the thermodynamic scalar curvature was proportional to the {\it correlation volume} of the system and this diverged at a critical point of second order phase transition. This direct connection between the thermodynamic scalar curvature and the microscopic correlation length make this geometrical framework extremely suitable for the description of black hole thermodynamics where a complete microscopic structure is till an elusive issue. In fact the application of this framework has provided interesting insights into the thermodynamics of both non extremal and extremal black holes \cite {Gibbons:1996af, Ferrara:1997tw, Sarkar:2008ji, Ruppeiner:2007up, Cai:1997nb, Cai:1998ep, Aman:2003ug, Aman:2006kd, Gibbons:2004ai, Shen:2005nu, Banerjee:2011cz, Banerjee:2012zm}.

Although the geometrical framework described characterized second order phase transitions through the divergence of the thermodynamic scalar curvature, a similar characterization based on the thermodynamic curvature, for first order phase transitions was lacking. One of the authors (GS) in the collaborations \cite{Sahay:2010wi,Sahay:2010tx,Sahay:2010yq} showed that the thermodynamic scalar curvature exhibited a multiple valued branched structure similar to the free energy at a first order phase transition which could be used for their characterization. Application of this to the study of liquid gas like first order phase transitions for a large class of asymptotically AdS black holes provided a confirmation of this alternative characterization. Following this in \cite {Ruppeiner:2011gm} a complete alternative geometrical characterization for first order phase transition, phase coexistence and supercritical phenomena could be established in the framework of thermodynamic geometries. This characterization extended Widoms \cite {Widom1696618} microscopic approach based on the correlation length to describe first order phase transitions,  to propose the equality of the correlation length in the coexisting phases. In the framework of thermodynamic geometries the above characterization translates to the equality of the thermodynamic scalar curvature $R$ for the coexisting phases. This essentially amounted to the crossing of the branches of the multiple valued curvature $R$ for the coexisting phases. This $R$- Crossing Formula \cite{Ruppeiner:2011gm} applied to conventional simple fluids as an alternative to the Maxwell equal area construction or the equality of the free energy, showed remarkable correspondence with experimental data . Additionally this framework \cite{Ruppeiner:2011gm} also provided a first theoretical method for the construction of the Widom line, the locus of the maxima of the correlation length in the supercritical regime. The Widom line serves as a line of dynamical crossover for fluid properties that seems to retain the memory of the distinct subcritical phases \cite {Widom1696618,Widom1974107,RevModPhys.71.S358,Simeoni:2010}. Thus our construction led to a complete unified geometrical framework for the characterization of subcritical, critical and supercritical phenomena based on the thermodynamic scalar curvature and has led to interesting further applications \cite {Dey:2011cs,PhysRevE.86.021130,PhysRevE.85.031201,PhysRevE.88.032123}.

Following earlier studies of the thermodynamics of electrically charged AdS black holes \cite{Chamblin:1999tk,Chamblin:1999hg, Kubiznak:2012wp} in \cite{Dutta:2013dca} the authors have investigated the thermodynamics of  dyonic charged AdS black holes \cite{Lu:2013ura} in four dimensions which involves both electric and magnetic charges. These dyonic charged black holes are solutions to the equations of motion of a Einstein-Maxwell theory with a negative cosmological constant. In \cite{Dutta:2013dca} the cosmological constant $\Lambda$ is considered as the thermodynamic pressure \cite{Henneaux:1984ji} and it was shown that in a mixed canonical-grand canonical ensemble with a fixed magnetic charge and a varying electric charge the black hole undergoes a liquid gas like first order phase transition culminating in a critical point that resembles the phase diagram of a Van der Waals fluid \cite{Banerjee:2011raa}. A similar result is also expected to follow for the fixed electric charge and varying magnetic charge scenario due to the symmetrical fashion in which the charges occur in the Smarr formula for the black hole mass. As these black holes exhibit first order phase transitions in mixed ensembles it is naturally interesting to study the phase structure using the thermodynamics scalar curvature $R$ referred earlier in the framework of thermodynamic geometries. In this article we investigate the thermodynamic geometry of such dyonic charged AdS black holes in four dimensions using mixed canonical grand canonical ensembles and study their phase transition and critical phenomena. However in our analysis we adopt a more conventional thermodynamic approach without considering the cosmological constant as a thermodynamic pressure. A study of the thermodynamics of this simpler case reproduces the first order phase transition and critical phenomena in a mixed ensemble observed in \cite{Dutta:2013dca} where the cosmological constant is considered as the thermodynamic pressure. In particular we implement our $R$- Crossing Formula for the thermodynamic scalar curvature $R$ to characterize the phase coexistence in the first order liquid gas like phase transition and show that the curvature $R$ diverges at the critical point in accordance with the conclusions in \cite{Sahay:2010wi,Sahay:2010tx,Sahay:2010yq,Ruppeiner:2011gm}.

This article is organized as follows, in Section 2 we briefly review the four dimnsional dyonic charged AdS black holes as solutions to the equation of motion of a Einsteic Maxwell AdS action. In Section 3 we discuss the thermodynamics of these dyonic charged AdS black holes in a more conventional scenario without considering the cosmological constant as a thermodynamic pressure and reproduce the first order phase transition and critical phenomena described in \cite {Dutta:2013dca}. In Section 4 we briefly review the essential elements of thermodynamic geometries and employ this geometrical framework to investigate the thermodynamics and phase transitions of the four dimensional dyonic charged AdS black holes and implement the $R$ crossing Method to characterize the first order liquid gas like phase transition for these black holes. In Section 5 we present a summary of our results and discussions including future open issues.

\section{Dyonic black holes}

In this section we briefly review the dyonic charged black hole solutions to the equations of motion of the Einstein Maxwell theory
in a four dimensional AdS space time as described in \cite {Dutta:2013dca}. The action for this theory may be expressed as 
\begin{equation}
I= \frac{1}{16\pi} \int{d^4x \sqrt{-g} (\ -R-\frac{6}{l^2}+\frac{1}{4}F_{\mu\nu}F^{\mu\nu})},\label{eq:action}
\end{equation}
where $R$ is the scalar curvature, $F_{\mu\nu}$ is the electromagnetic field strength tensor and $l$ is the AdS length scale and the gravitational constant $G$ has been set equal to one.
The equations of motions may be written down as
\begin{eqnarray}
R_{\mu\nu}-\frac{1}{2}g_{\mu\nu}\left(R+\frac{3}{l^2}\right) &=& 2(F_{\mu\lambda}F_{\nu}^{~\lambda}-\frac{1}{4}F_{\alpha\beta}F^{\alpha\beta}),\label{eq:EinsEom}\\
\nabla_{\mu}F^{\mu\nu} &=& 0. \label{eq:MaxwEom}
\end{eqnarray}
A static spherically symmetric solution to the equations of motion are
\begin{eqnarray}
ds^2 = - f(r) dt^2 + \frac{dr^2}{f(r)} + r^2 d\theta^2 + r^2 \sin^2\theta d\phi^2,\label{eq:metric}\\
f(r) = 1+ \frac{r^2}{l^2} - \frac{2M}{r} + \frac{q_{e}^2+q_{m}^2}{r^2},\label{eq:lapseFunc}\\
A_{\mu}dx^{\mu} = q_{e}\left(\frac{1}{r_{+}}-\frac{1}{r}\right)dt+ (q_{m}\cos\theta)d\phi,\label{eq:GaugeField}
\end{eqnarray}
where, $A_{\mu},q_{e},q_{m}$ and $M$ are identified as the electromagnetic four-potential, electric charge, magnetic charge
and mass of the black hole respectively. The quantity $r_{+}$ is the outer horizon radius of the black hole given by the zeroes of the lapse function $f(r)$ described above.

A dyonic black hole is associated with a dyonic charge that involves both an electric ($ q_e $) and a magnetic ($ q_m $) charge \cite{Kubiznak:2012wp,Dutta:2013dca}. The presence of magnetic charge leads to a corresponding magnetic potential ($ \phi_m $) in addition to the electric potential ($\phi_e$). The expression for  electric potential ($\phi_e$) may be given as,
\begin{equation}
\phi_e = \phi_E-\frac{q_e}{r}=q_e(\frac{1}{r_+}-\frac{1}{r}),
\end{equation}
here, $\phi_{E}$ may be defined as  the asymptotic value of the electric potential measured at infinity.

\section{Thermodynamics of Dyonic charged black holes}

It is now well established that in a semi classical framework, black holes may be considered as thermodynamic systems characterized by a Hawking temperature $T$ and an entropy $S$. For the case of the four dimensional dyonic charged AdS black hole these may be expressed as
\begin{eqnarray}
T &=& \frac{1}{4\pi r_{+}}\left[1+\frac{3 r_{+}^2}{l^2}-\frac{q_{e}^2}{r_{+}^2}-\frac{q_{m}^2}{r_{+}^2}\right],\label{eq:DyonicT}\\
S &=& \pi r_{+}^2.\label{eq:entropyr}
\end{eqnarray}

The first law of black hole thermodynamics in this case may be written down as
\begin{equation}
dM = Tds+\phi_e dq_e+\phi_m dq_m,\label{eq:firstlaw}
\end{equation}
where, $T$ and $S$ are the Hawking temperature and the entropy of the dyonic charged AdS black hole respectively. Using this the Smarr formula for the mass $M$ of the black hole in terms of $ s , q_{e}$ and $q_{m}$ using Eq.(\ref{eq:lapseFunc}) 
may be obtained as in \cite{Dutta:2013dca} 
\begin{equation}
M =\frac{1}{2} \sqrt{\frac{\pi}{s}}\  \bigg(\frac{s^2}{\pi^2} + \frac{s}{\pi} + q_{e}^2+q_{m}^2\ \bigg).\label{eq:smarr}
\end{equation}

Following \cite{Sahay:2010wi} we have scaled out the AdS length scale $ l $ which do not figure in our subsequent analysis. To analyze the thermodynamics of the dyonic charged AdS black hole we consider a mixed ensemble that is grand canonical with respect to the varying electric charge $ \phi_e $ but canonical with respect to the magentic charge $ q_m $ which is fixed. Since both the magnetic as well as the electric charges enter in the Smarr relation given by Eq.(\ref{eq:smarr}) in a symmetric fashion, hence the alternative mixed ensemble with a varying magnetic charge $ q_m $ and fixed electric charge $ q_e $ is expected to have an identical thermodynamic behavior and does not need to be considered seprately. Using the Smarr formula the Hawking temperature $ T $ may be expressed as
\begin{equation}
T= \frac{dM}{ds}= \frac{1}{4\pi^{\frac{3}{2}}} \frac{1}{s^\frac{3}{2}} [3s^2 + \pi s - \pi^2 (q_{e}^2+q_{m}^2)].\label{eq:SmarrT}
\end{equation}

The Gibbs free energy specific to the mixed ensemble involving a varying electric charge $ q_e $ and a fixed magnetic charge $ q_m $ is as follows
\begin{equation}
G= M - T s - \phi_e q_e\label{eq:FreeEn1}
\end{equation}
where $\phi_e$ is the fixed electric potential.

It is now possible to express the Gibbs free energy $G$ given above in terms of the variables $ \phi_e, q_m $ and $ S $ as
\begin{equation}
G= \frac{1}{4\pi^2} \sqrt{\frac{\pi}{s}} \ [3\pi^2 q_{m}^2 - S^2 + \pi S (1-\phi_{e}^2)\ ].\label{eq:FreeEn2}
\end{equation}

Similarly the temperature of the dyonic charged AdS black hole may now be expressed in terms of the variables $S, \phi_{e}$ and $q_{m}$ as
\begin{equation}
T=\frac{ \left(-\pi^2 q_{m}^2+3 s^2+\pi  \left(s-s\phi_{e}^2\right)\right)}{4 (\pi s)^{3/2}}.\label{eq:SmarrPhiT}
\end{equation}

\begin{figure}[H]
\begin{center}
\includegraphics[width =2.8in,height=1.8in]{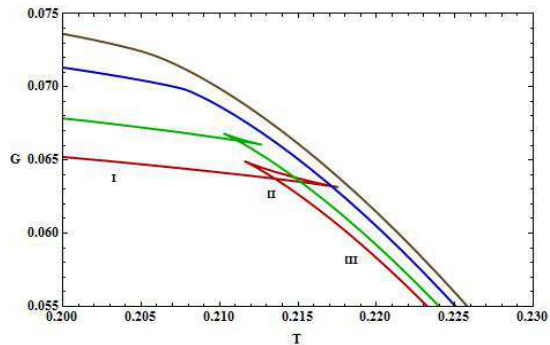}
\caption{\label{fig:GvsT} Plots of $ G $ with $ T $. The red, green, blue and the brown curves correspond to $(\phi_e = 0.6, q_m=0.095)$ , $(\phi_e = 0.6, q_m=0.1)$ , $(\phi_e = 0.6, q_m=0.107)$ and $(\phi_e = 0.6, q_m=0.13)$ respectively.}
\end{center}
\end{figure}

Now using these equations (\ref{eq:FreeEn2}) and (\ref{eq:SmarrPhiT}) for the Gibb' s free energy and the Hawking temperature 
it is straightforward to obtain the variation of  the Gibb's free energy $G$ with respect to the temperature $ T $ as shown in Fig.(\ref{fig:GvsT}). The plot for the free energy $G$ shows a typical multiple valued branched {\it swallowtail} behavior characteristic of a first order phase transition at a certain temperature $T$.  This behavior disappears as the value of the magnetic charge $q_m$ is increased demonstrating the culmination of the subcritical phase coexistence to a critical point of a second order phase transition. This  may be observed  from  the Fig.(\ref{fig:GvsT}) where the blue curve corresponds to the free energy near the critical point. The red curve in Fig.(\ref{fig:GvsT}) corresponds to the value $q_m =0.095$ for the magnetic charge which is below the critical value $q_m=0.107$. The subcritical red curve possesses three distinct branches that form the typical {\it swallowtail} structure for the fixed value of the magnetic charge $q_m=0.095$ whereas above the critical value $ q_m = 0.107 $ this structure disappears. It may be noted that the qualitative nature of the plots do not change with the variation of electric potential $ \phi_e $.

\begin{figure}[H]
\begin{center}
\includegraphics[width=2.8in,height=1.8in]{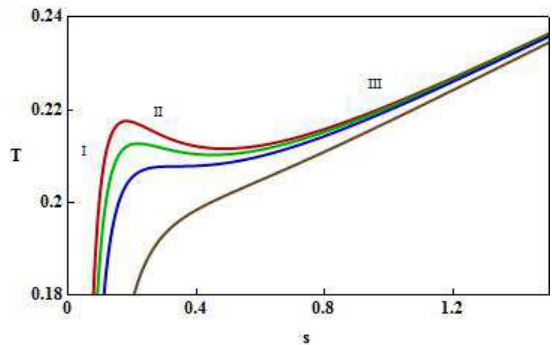}
\caption{\label{fig:TvsS} $T$ vs. $S$ plot for different values of $q_m$. The red, green, blue and the brown curves correspond to values of $q_m=0.095$, $q_m=0.1$, $q_m=0.107$ and $q_m=0.13$ respectively.} 
\end{center}
\end{figure} 

The first order phase transition described here occurs between the two distinct phases typified by small and large black holes characterized by their (outer) horizon radius $r_{+}$, as  shown in the Fig.(\ref{fig:TvsS}),  where the temperature $T$ is plotted against the entropy $S$. Branch I or the `small black hole' branch here corresponds to the `liquid-like' phase and branch III or the `large black hole' branch corresponds to the `gas-like' phase in relation to a the usual liquid-gas phase transition. The three branches of the red curve in the $T-S$ plot labeled as $I,II$ and $III$ correspond to the three respective branches of the red curve in the $G-T$ plot displayed in Fig.(\ref{fig:GvsT}). As one increases the temperature $ T $ the system moves along the branch-I, reaches the point of crossing of the branch-I and branch-III, jumps to the branch-III and continues along that branch. Branch-II is least favored  as it corresponds to a higher free energy $G$ than the other two branches at any given temperature $T$. As is well known this branch corresponds to a meta-stable state where the specific heat becomes negative. As one proceeds beyond the critical point, branch-II disappears and this corresponds to the usual {\it uniform supercritical phase}. Notice that in the limit $ q_m \rightarrow 0$, the resulting black hole is essentially a conventional electrically charged RN-AdS black hole in grand-canonical (fixed $ \phi_e $ ensemble). In this limit the only possible  phase transition is the usual  Hawking-Page transition \cite{Hawking:1982dh,Page:2004xp} between a thermal AdS space time and a black hole solution. In this limit the small black hole branch coincides with the $ T $-axis in $ G-T $ plot.

\section{Thermodynamic Geometry of  a Dyonic Charged Black Hole}

In this section we briefly collect the relevant results of thermodynamic geometry and then procced to obtain the thermodynamic geometry and the scalar curvature of the dyonic charged black hole in the mixed canonical-grand canonical ensemble elucidated earlier. As mentioned in the introduction it is possible to associate a curved Riemannian geometry with an Euclidean signature based on the fluctuations of the thermodynamic variables in the equlibrium state space of the system. The positive definite invariant line element for this geometry is then related to the probability distribution of the fluctuations connecting such equlibrium thermodynamic states in a Gaussian approximation. The Riemannian metric describing this geometry may be obtained from the thermodynamic potential appropriate to the representation being considered. In the entropy representation \cite{Ruppeiner:2011gm}, the metric is given by the Hessian of the total entropy $S= S(U,V, N)$ with respect to the extensive variables with the volume $V$ held fixed and $U,N$ being the
the internal energy and the number density respectively. The metric may be represented as
\begin{equation}
g_{\mu\nu}=-\frac{\partial^2 s}{\partial x^\mu \partial x^\nu},\label{eq:Rwe shallupp}
\end{equation}
where $ {x^\mu} $ are the extensive variables which serve as the coordinates in the state space subject to the condition $dS\geq 0$ in accordance with the second law of thermodynamics and $s$ is the combined entropy of the system and its surroundings. The volume $V$ is considered to be infinite in the thermodynamic limit and in this case $s$ may be considered to be just the entropy of the system only. Alternatively in the energy representation 
 \cite{Weinhold:1975hp, Weinhold:1975sc}, the thermodynamic metric is obtained from the Hessian of the internal energy $U$ with respect to the extensive variables and may be expressed as
\begin{equation}
g_{\mu\nu}=\frac{1}{T} \frac{\partial^2 U}{\partial x^\mu \partial x^\nu}.\label{eq:Wein}
\end{equation}

Using scaling and hyperscaling arguments the thermodynamic scalar curvature $ R $ obtained from this metric may be shown to be proportional to the correlation volume of the system as
\begin{equation}
R \sim \xi^d,\label{eq:Rcorr}
\end{equation}
where $ \xi $ is the correlation length and $ d $ is the physical dimension of the system. Thus the thermodynamic scalar curvature $R$ actually relates to the underlying microscopic interactions with $R=0$ describing a non interacting microscopic basis.

For our case of the four dimensional dyonic charged AdS black hole in the mixed canonical-grand canonical ensemble it is convenient to use the energy representation due to Weinhold \cite {Weinhold:1975hp} to obtain the thermodynamic metric. From the Smarr formula Eq.(\ref{eq:smarr}) the extensive variables which serve as the coordinates for the geometry are $ x^\mu =\{s, q_e\} $ the electric charge and the entropy respectively. Here the magnetic charge $ q_m $ is held fixed. The thermodynamic metric may then be expressed as
\begin{equation}\label{eq:19}
g_{\mu\nu}=\frac{1}{T} \frac{\partial^2 M}{\partial x^\mu \partial x^\nu}
\end{equation}
It is to be noted that the internal energy of the black hole is determined to be its ADM mass $M$ in the formulation for black hole thermodynamics which we are using.

The corresponding metric components in this representation with the coordinates $ x^\mu =\{s, q_e\} $ may be expressed as follows
\begin{eqnarray}
g_{s s} &=& \frac{3\pi^2\left(q_{e}^2+q_{m}^2\right)-\pi s+3 s^2}{2 s \left(-\pi ^2 \left(q_{e}^2+q_{e}^2\right)+\pi s+3 s^2\right)}, \nonumber \\
g_{s q_{e}}= g_{q_{e} s} &=& -\frac{2 \pi ^2 q_{e}}{-\pi ^2 \left(q_{e}^2+q_{m}^2\right)+\pi  s+3 s^2}, \nonumber \\
g_{q_{e} q_{e}} &=& \frac{4 \pi ^2 s}{-\pi ^2 \left(q_{e}^2+q_{m}^2\right)+\pi  s+3 s^2}.\label{eq:weinmetric}
\end{eqnarray}

From the metric in the energy representation in Eq.(\ref{eq:weinmetric}) using the preceeding formula the thermodynamic scalar curvature $R$ in terms of the variables $s, q_{e}$ and $q_{m}$ may be expressed  from the standard Riemannian geometry formulation as,
\makeatletter 
\def\@eqnnum{{\normalsize \normalcolor (\theequation)}} 
\makeatother
{\small
\begin{eqnarray}
R &=& - \frac{N}{D}, \nonumber\\
N &=& (\pi ^2 q_{m}^2+3 s^2)(3\pi^4(q_{e}^2+ q_{m}^2)^2+\pi^3(-3q_{e}^2+q_{m}^2)s \nonumber\\
&+& 12 \pi^2(q_{e}^2+3 q_{m}^2) s^2 - 9\pi s^3+9 s^4),\nonumber\\
D &=& s(\pi^2(q_{e}^2+3 q_{m}^2)-\pi s + 3 s^2)^2(-\pi^2(q_{e}^2+q_{m}^2)\nonumber \\
&+&\pi s + 3 s^2).\label{eq:ScalarRqe}
\end{eqnarray}
}

For our purpose of analyzing the phase structure of the four dimensional dyonic charged black hole it would be convenient to express the thermodynamic scalar curvature $R$ in terms of the alternative variables $s, \phi_{e}$ and $q_{m}$ as
\makeatletter 
\def\@eqnnum{{\normalsize \normalcolor (\theequation)}} 
\makeatother
{\small
\begin{eqnarray}
R &=& - \frac{A}{B}, \nonumber\\
A &=& (\pi^2 q_{m}^2+3 s^2)(3 \pi^4 q_{m}^4+9 s^4+3 \pi s^3 (-3+4 \phi_{e}^2)\nonumber\\
&+&\pi ^3 q_{m}^2 s(1+6\phi_{e}^2)+3 \pi ^2 s^2(12 q_{m}^2-\phi_{e}^2+\phi_{e}^4)),\nonumber\\
B &=& s(3\pi^2 q_{m}^2+3 s^2+\pi s(-1+\phi_{e}^2))^2(-\pi^2 q_{m}^2\nonumber\\
&+& 3 s^2+\pi(s-s \phi_{e}^2)).\label{eq:ScalarRphie}
\end{eqnarray}
}

\begin{figure}[H]
\begin{center}
\includegraphics[width=2.8in,height=1.8in]{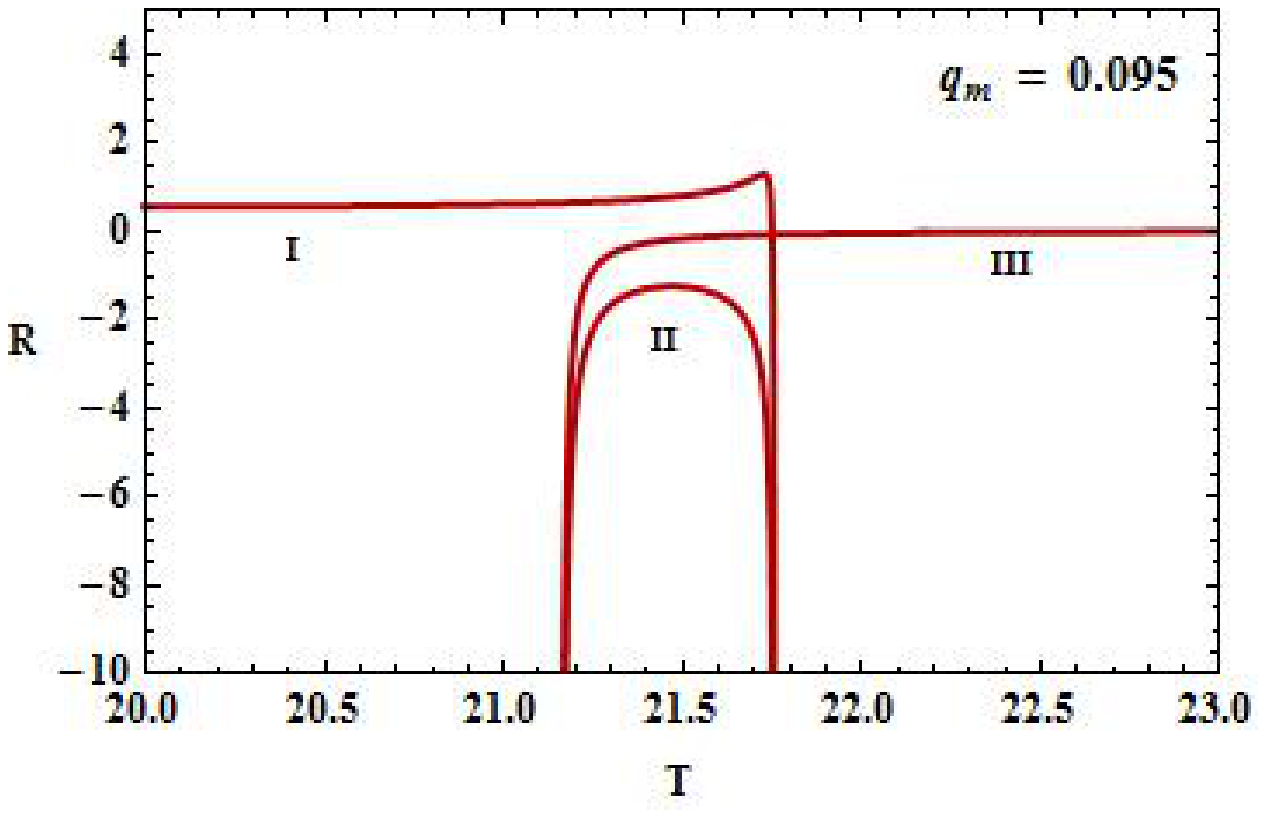}\\
\includegraphics[width=2.8in,height=1.8in]{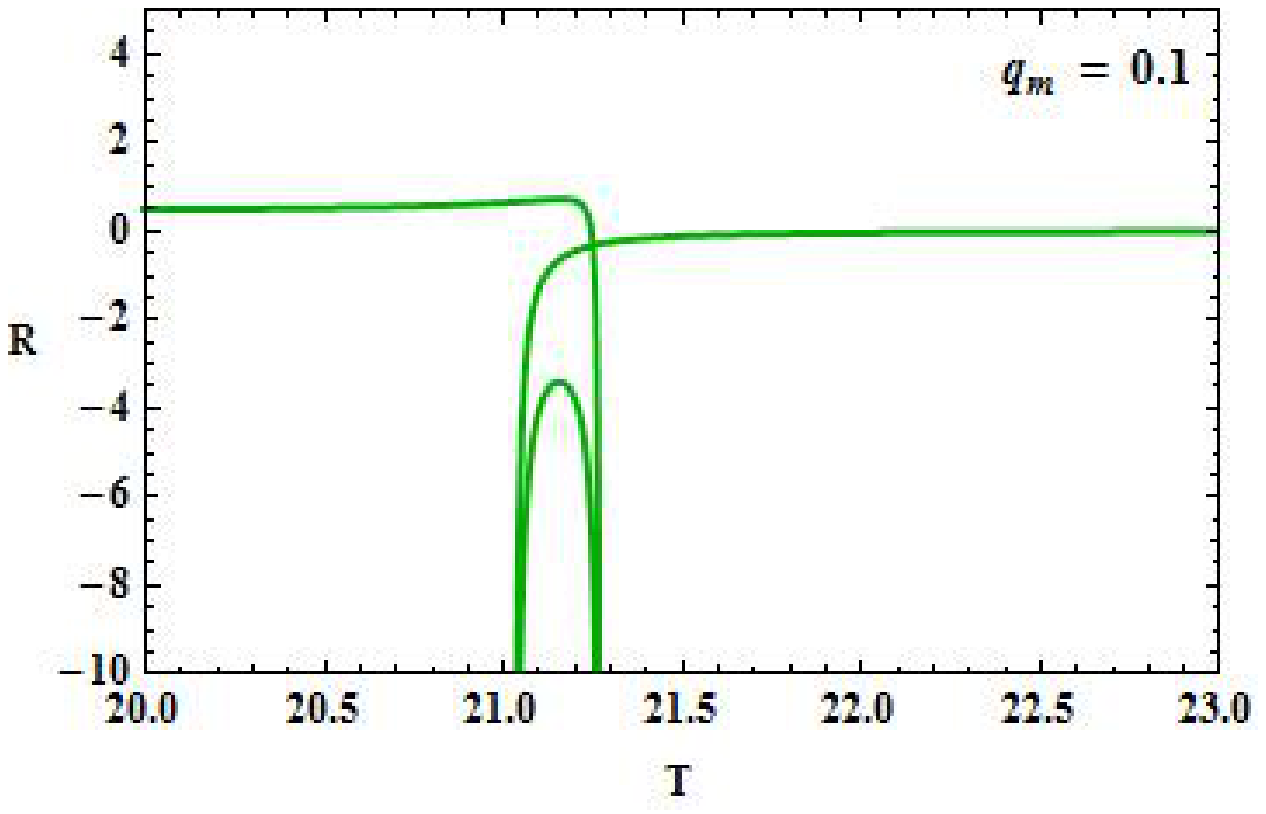}\\
\includegraphics[width=2.8in,height=1.8in]{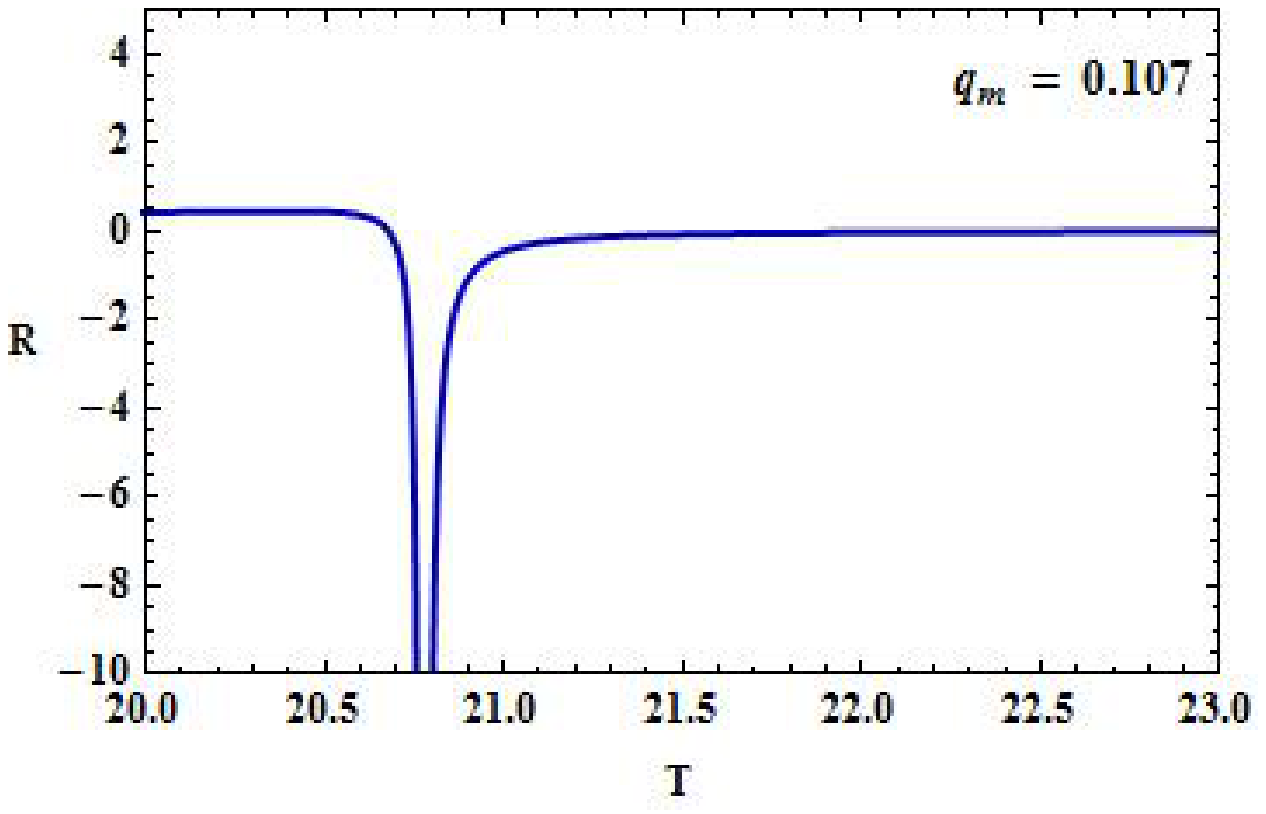}\\

\caption{\label{fig:RvsTqm} Plots of $ R $ with $ T $ for $ \phi_e = 0.6 $. The plots (a) and (b) with red and green curves are  for $ q_m =0.095 $ and $q_m = 0.1$ respectively while the blue curve in plot (c) is for critical value $ q_m = 0.107 $. In these plots, the $ R $ axes have been scaled down with a factor of $ 100 $ and the $ T $ axes are scaled up by the same factor to bring the important features of the plots within a suitable region.}
\end{center}
\end{figure}

The phase structure of the dyonic charged black hole may now be investigated by studying the variation of the thermodynamic scalar curvature $R$ with the temperature $ T $ using the equations (\ref{eq:ScalarRphie}) and (\ref{eq:SmarrPhiT})\footnote{ Notice that
in these plots, the  $ R $ axes have been scaled down by a factor of $ 100 $ and the $ T $ axes are scaled up by the same factor to present the important features within a small region.}. A comparison of the plots for the free energy of the dyonic charged AdS black hole in Fig.(\ref{fig:GvsT}) and those for the thermodynamic scalar curvature $ R $ with the temperature $ T $ in Fig.(\ref{fig:RvsTqm}) clearly shows the correspondence between the two. As mentioned earlier the thermodynamic scalar curvature $R$ is a multivalued function of the thermodynamic parameters in the neighbourhood of a first order phase transition simmilar to the free energy $G$. The red curve in the Fig.(\ref{fig:GvsT}) for the value of the magnetic charge $ q_m=0.095 $ corresponds to the subcritical regime accordingly has three distinct branches Moving in the direction of the increasing temperature $T$ 
we first encounter the branch on the left that corresponds to the `small black hole' phase (I). As the temperature $ T $ is increased, the system moves towards the right in the graph to reach the first order phase transition temperature at $ T=0.2137 $ where the `small black hole' branch (I) and the `large black hole' branch (III) intersect each other forming a `swallowtail'. At this temperature, the black hole makes a transition from the `small black hole' phase to the `large black hole' phase. As the temperature is further increased the system continues along the branch corresponding to the `large black hole' phase. 

Now from the red curve for the value of the magnetic charge $q_m=0.095$ shown in the Fig.(\ref{fig:RvsTqm}a) for the $R$ vs. $T$ plot a behaviour simmilar to that of the red curve in free energy $G$ vs the temperature $T$ plot described above may be observed. In this plot also, in the Fig.(\ref{fig:RvsTqm}a) 
moving in the direction of the increasing temperature $T$ the branch on the left that corresponds to the `small black hole' pahse (I) is first encountered. As the temperature $ T $ is further increased, the system moves towards the right to reach the first order phase transition temperature at $ T=0.2174 $ where the `small black hole' branch (I) and the `large black hole' branch (III) intersect each other.  For higher temperatures the system continues along the branch corresponding to the`large black hole' phase. In table (\ref{table:Tc}) we summarize the values of the first order phase transition temperatures $T_f$ obtained from the free energy $G$ vs. the temperature $T$ plots in Fig.(\ref{fig:GvsT}) and the thermodynamic scalar curvature $R$ vs $T$ plots in the Fig.(\ref{fig:RvsTqm}). It may be noted from the preceeding table that the difference in the values of the temperature $T_f$ for the first order phase transition obtained from both the $G-T$ plots and the $R-T$ plots respectively matches upto the second decimal place. At this point it must be emphasized that the first order phase transition temperature obtained from the free energy $G$ vs the temperature $T$ plot essentially corresponds to the Maxwell construction signifying equal free energy in the coexisting phases. Whereas the corresponding temperature obtained from the thermodynamic scalar curvature $R$ vs the temperature $T$ plot through the crossing of the distinct branches is characterized by the $R$ Crossing Method proposed by us as an alternative to the Maxwell construction outlined in \cite{Sahay:2010wi,Sahay:2010tx,Ruppeiner:2011gm}. This essentially corresponds to the equality of the correlation length $\xi$ in the coexisting phases proposed as an extension of  Widom's microscopic approach to phase transitions. It also must be emphasized here that unlike the subject of black hole thermodynamics where no experimental results are available for comparison our alternative $R$ Crossing Method for characterization of first order phase transition exhibits a better fit with experimental data for simple fluids as described in \cite{Ruppeiner:2011gm}.

\begin{table}[H]
\centering
\begin{tabular}{|c|c|c|c|}
\hline 
\textbf{$\phi_{e} = 0.6 $} & \textbf{$q_m=0.095$} & \textbf{$q_m=0.1$} & \textbf{$q_m=0.107$}\tabularnewline
\hline
\textbf{\small $T_f$(G vs T)} & 0.2137 & 0.2115 & 0.2077 \tabularnewline
\hline
\textbf{\small $T_f$(R vs T)} & 0.2174 & 0.2125 & 0.2077 \tabularnewline
\hline
\end{tabular}
\caption{\label{table:Tc} Table showing the values of the temperature $T_f$ for the first order phase transition obtained form both the $G-T$ plots and the $R-T$ plots for a fixed value of $\phi_e =0.6$ and varying values of $q_m$.}
\end{table}

Further observe now from the plots for  the thermodynamic scalar curvature $ R$ vs temperature $T$  that upon increasing the value of the magnetic charge $ q_m $, the `small black hole' branch (I) and the `large black hole' branch (III) recede away from each other as one approaches criticality. At the critical point the crossing of the two branches of the thermodynamic scalar curvature $R$ completely disappear resulting in a divergence of $R$ as predicted by Eq.(\ref{eq:Rcorr}). The blue curve in the Fig.(\ref{fig:RvsTqm}) corresponds to the critical point of  phase transition which clearly show that both the branches of the thermodynamic scalar curvature extend to the infinity. 
It may be observed from the Table. (\ref{table:Tc}) that the thermodynamic scalar curvature diverges at a critical temperature
which matches well with the critical temperature obtained from the free energy $G$ vs temperature $T$ plot (\ref{fig:GvsT}).
In other words the critical value of the temperature obtained from the thermodynamic scalar curvature $R$ vs the temperature $T$ plot in Fig.(\ref{fig:RvsTqm}c) almost coincides with the value of the critical temperature obtained from the free energy $G$ vs the temperature $ T $ plot ( the blue curve in Fig.(\ref{fig:GvsT})). Beyond the critical point the distinct subcritical phases of the {\it small} and {\it large black holes} disappear resulting in a {\it uniform supercritical phase} exactly similar to the case of a liquid gas phase transition. Notice however that for the case of fluids as described in \cite{Ruppeiner:2011gm}, the fluid retains the memory of the subcritical phases across a locus of the maxima of the correlation length which is termed the Widom line in the literature, which serves as a crossover line for dynamical fluid properties and divides the supercritical regime into a liquid like and gas like region. The significance of such a Widom line in black hole thermodynamics however is not quite clear although just as in \cite{Ruppeiner:2011gm} it is possible to construct the Widom line through the locus of the maxima of the thermodynamic scalar curvature $R$ in the supercritical regime.

\section{Summary and Discussions}

In summary we have investigated the thermodynamics, phase transition and critical phenomena of four dimensional dyonic charged
AdS black holes using the framework of thermodynamic geometries. Unlike \cite {Dutta:2013dca} where the cosmological constant is considered as the thermodynamic pressure in our study we have adopted a more conventional approach where the cosmological constant is a fixed parameter for the black hole solution. In this context we have reproduced the first order liquid gas like phase transition with coexisting phases in this more simple scenario using a mixed canonical grand canonical ensemble where the magnetic charge is held fixed and the electric charge is allowed to vary. Due to the symmetrical appearence of the electric and the magnetic charge in the Smarr formula, in the alternative setting of a fixed electric charge and varying magnetic charge the thermodynamic behaviour is expected to be identical and does not merit a separate analysis. Our analysis in this simpler scenario for the Gibb' s free energy as a function of the temperature clearly reproduces the typical {\it swallowtail} feature characterizing the first order liquid gas like phase transition and coexistence between a small and a large black hole phase in the subcritical regime. 

The thermodynamic metric for the dyonic charged black hole is then obtained in the more convenient Weinhold picture as a Hessian of the internal energy with respect to the entropy $S$ and the electric charge $\phi_e$. The thermodynamic scalar curvature $R$ for this black hole is then obtained using standard Riemannian geometric techniques for the Ricci scalar. Analysis of the variation of the thermodynamic scalar curvature $R$ as a function of the temperature $T$ clearly demonstrates the expected branching behaviour signifying a first order phase transition.  The plot of $R$ vs the temperature $T$ exhibits the crossing of the branches at the first order phase transition temperature and the fixed value of the magnetic charge $\phi_m$ in accordance with our {\it $R$ Crossing Method} that serves as an alternative to the conventional Maxwell construction. The value of the magnetic charge for the first order phase transition obtained through the {\it $R$ Crossing Method} compares favorably and is almost identical to that obtained from the usual Maxwell construction and the equality of the Gibbs free energy $G$ for the coexisting phases. For a further increase of the magnetic charge away from the subcritical regime for the first order phase transition values finally leads to a critical point signifying a second order phase transition where the thermodynamic scalar curvature $R$ exhibits an expected divergence. The branching and subsequent crossing behaviour leading to a divergence at the critical point of the thermodynamic scalar curvature $R$ is completely simmilar to that for the van der Waals gas described in \cite{Sahay:2010wi, SahayThesis:2010}. 

Our analysis clearly establishes and further corroborates our approach for studying the phase structure of AdS black holes in the framework of thermodynamic geometry and proves our proposal for the $R$ Crossing Method as an alternative to the conventional Maxwell construction and equality of the Gibbs free energy required for phase coexistence. An important open problem for future investigation in this scenario is to implement the framework of thermodynamic geometry to describe {\it extended black hole thermodynamics} which includes the cosmological constant as a thermodynamic pressure. In addition it would be interesting to construct the Widom line in the supercritical regime through the locus of the maxima of the thermodynamic scalar curvature $R_{max}$ as suggested by us and investigate the significance of this Widom line in the context of black hole phase transitions.

\section{Acknowledgment}
This work of Pankaj Chaturvedi is supported by grant no. 09/092(0846)/2012-EMR-I from CSIR India.

\bibliography{DyonicBHpaper}

\begin{thebibliography}{10}

\bibitem{Hawking:1974sw}
S.W. Hawking.
\newblock {Particle Creation by Black Holes}.
\newblock {\em Commun.Math.Phys.}, 43:199--220, 1975.

\bibitem{Bekenstein:1973ur}
Jacob~D. Bekenstein.
\newblock {Black holes and entropy}.
\newblock {\em Phys.Rev.}, D7:2333--2346, 1973.

\bibitem{Bardeen:1973gs}
James~M. Bardeen, B.~Carter, and S.W. Hawking.
\newblock {The Four laws of black hole mechanics}.
\newblock {\em Commun.Math.Phys.}, 31:161--170, 1973.

\bibitem{Wald:1999vt}
Robert~M. Wald.
\newblock {The thermodynamics of black holes}.
\newblock {\em Living Rev.Rel.}, 4:6, 2001.

\bibitem{Page:2004xp}
Don~N. Page.
\newblock {Hawking radiation and black hole thermodynamics}.
\newblock {\em New J.Phys.}, 7:203, 2005.

\bibitem{Ross:2005sc}
Simon~F. Ross.
\newblock {Black hole thermodynamics}.
\newblock 2005.

\bibitem{Townsend:1997ku}
P.K. Townsend.
\newblock {Black holes: Lecture notes}.
\newblock 1997.

\bibitem{Dabholkar:2012zz}
Atish Dabholkar and Suresh Nampuri.
\newblock {Quantum black holes}.
\newblock {\em Lect.Notes Phys.}, 851:165--232, 2012.

\bibitem{Hawking:1982dh}
S.W. Hawking and Don~N. Page.
\newblock {Thermodynamics of Black Holes in anti-De Sitter Space}.
\newblock {\em Commun.Math.Phys.}, 87:577, 1983.

\bibitem{Chamblin:1999tk}
Andrew Chamblin, Roberto Emparan, Clifford~V. Johnson, and Robert~C. Myers.
\newblock {Charged AdS black holes and catastrophic holography}.
\newblock {\em Phys.Rev.}, D60:064018, 1999.

\bibitem{Chamblin:1999hg}
Andrew Chamblin, Roberto Emparan, Clifford~V. Johnson, and Robert~C. Myers.
\newblock {Holography, thermodynamics and fluctuations of charged AdS black
  holes}.
\newblock {\em Phys.Rev.}, D60:104026, 1999.

\bibitem{Caldarelli:1999xj}
Marco~M. Caldarelli, Guido Cognola, and Dietmar Klemm.
\newblock {Thermodynamics of Kerr-Newman-AdS black holes and conformal field
  theories}.
\newblock {\em Class.Quant.Grav.}, 17:399--420, 2000.

\bibitem{Kubiznak:2012wp}
David Kubiznak and Robert~B. Mann.
\newblock {P-V criticality of charged AdS black holes}.
\newblock {\em JHEP}, 1207:033, 2012.

\bibitem{Weinhold:1975sc}
F.~Weinhold.
\newblock Metric geometry of equilibrium thermodynamics.
\newblock {\em The Journal of Chemical Physics}, 63(6), 1975.

\bibitem{Weinhold:1975hp}
F.~Weinhold.
\newblock Metric geometry of equilibrium thermodynamics. ii. scaling,
  homogeneity, and generalized gibbs–duhem relations.
\newblock {\em The Journal of Chemical Physics}, 63(6), 1975.

\bibitem{Ruppeiner:1995zz}
George Ruppeiner.
\newblock {Riemannian geometry in thermodynamic fluctuation theory}.
\newblock {\em Rev.Mod.Phys.}, 67:605--659, 1995.

\bibitem{Gibbons:1996af}
Gary~W. Gibbons, Renata Kallosh, and Barak Kol.
\newblock {Moduli, scalar charges, and the first law of black hole
  thermodynamics}.
\newblock {\em Phys.Rev.Lett.}, 77:4992--4995, 1996.

\bibitem{Ferrara:1997tw}
Sergio Ferrara, Gary~W. Gibbons, and Renata Kallosh.
\newblock {Black holes and critical points in moduli space}.
\newblock {\em Nucl.Phys.}, B500:75--93, 1997.

\bibitem{Sarkar:2008ji}
Tapobrata Sarkar, Gautam Sengupta, and Bhupendra~Nath Tiwari.
\newblock {Thermodynamic Geometry and Extremal Black Holes in String Theory}.
\newblock {\em JHEP}, 0810:076, 2008.

\bibitem{Ruppeiner:2007up}
George Ruppeiner.
\newblock {Black Holes: Fermions at the Extremal Limit?}
\newblock 2007.

\bibitem{Cai:1997nb}
Rong-Gen Cai.
\newblock {Critical behavior in black hole thermodynamics}.
\newblock {\em J.Korean Phys.Soc.}, 33:S477--S482, 1998.

\bibitem{Cai:1998ep}
Rong-Gen Cai and Jin-Ho Cho.
\newblock {Thermodynamic curvature of the BTZ black hole}.
\newblock {\em Phys.Rev.}, D60:067502, 1999.

\bibitem{Aman:2003ug}
Jan~E. Aman, Ingemar Bengtsson, and Narit Pidokrajt.
\newblock {Geometry of black hole thermodynamics}.
\newblock {\em Gen.Rel.Grav.}, 35:1733, 2003.

\bibitem{Aman:2006kd}
Jan~E. Aman, Ingemar Bengtsson, and Narit Pidokrajt.
\newblock {Flat information geometries in black hole thermodynamics}.
\newblock {\em Gen.Rel.Grav.}, 38:1305--1315, 2006.

\bibitem{Gibbons:2004ai}
G.W. Gibbons, M.J. Perry, and C.N. Pope.
\newblock {The First law of thermodynamics for Kerr-anti-de Sitter black
  holes}.
\newblock {\em Class.Quant.Grav.}, 22:1503--1526, 2005.

\bibitem{Shen:2005nu}
Jian-yong Shen, Rong-Gen Cai, Bin Wang, and Ru-Keng Su.
\newblock {Thermodynamic geometry and critical behavior of black holes}.
\newblock {\em Int.J.Mod.Phys.}, A22:11--27, 2007.

\bibitem{Banerjee:2011cz}
Rabin Banerjee and Dibakar Roychowdhury.
\newblock {Critical phenomena in Born-Infeld AdS black holes}.
\newblock {\em Phys.Rev.}, D85:044040, 2012.

\bibitem{Banerjee:2012zm}
Rabin Banerjee and Dibakar Roychowdhury.
\newblock {Critical behavior of Born Infeld AdS black holes in higher
  dimensions}.
\newblock {\em Phys.Rev.}, D85:104043, 2012.

\bibitem{Sahay:2010wi}
Anurag Sahay, Tapobrata Sarkar, and Gautam Sengupta.
\newblock {Thermodynamic Geometry and Phase Transitions in Kerr-Newman-AdS
  Black Holes}.
\newblock {\em JHEP}, 1004:118, 2010.

\bibitem{Sahay:2010tx}
Anurag Sahay, Tapobrata Sarkar, and Gautam Sengupta.
\newblock {On the Thermodynamic Geometry and Critical Phenomena of AdS Black
  Holes}.
\newblock {\em JHEP}, 1007:082, 2010.

\bibitem{Sahay:2010yq}
Anurag Sahay, Tapobrata Sarkar, and Gautam Sengupta.
\newblock {On The Phase Structure and Thermodynamic Geometry of R-Charged Black
  Holes}.
\newblock {\em JHEP}, 1011:125, 2010.

\bibitem{Ruppeiner:2011gm}
George Ruppeiner, Anurag Sahay, Tapobrata Sarkar, and Gautam Sengupta.
\newblock {Thermodynamic Geometry, Phase Transitions, and the Widom Line}.
\newblock {\em Phys.Rev.}, E86:052103, 2012.

\bibitem{Widom1696618}
B.~Widom.
\newblock Equation of state in the neighborhood of the critical point.
\newblock {\em The Journal of Chemical Physics}, 43(11), 1965.

\bibitem{Widom1974107}
B.~Widom.
\newblock The critical point and scaling theory.
\newblock {\em Physica}, 73(1):107 -- 118, 1974.

\bibitem{RevModPhys.71.S358}
H.~Eugene Stanley.
\newblock Scaling, universality, and renormalization: Three pillars of modern
  critical phenomena.
\newblock {\em Rev. Mod. Phys.}, 71:S358--S366, Mar 1999.

\bibitem{Simeoni:2010}
G.~G. Simeoni, T.~Bryk, F.~A. Gorelli, M.~Krisch, G.~Ruocco, M.~Santoro, and
  T.~Scopigno.
\newblock { The Widom line as the crossover between liquid-like and gas-like
  behaviour in supercritical fluids}.
\newblock {\em Nature Physics}, 6:503--507, 2010.

\bibitem{Dey:2011cs}
Anshuman Dey, Pratim Roy, and Tapobrata Sarkar.
\newblock {Information geometry, phase transitions, and the Widom line:
  Magnetic and liquid systems}.
\newblock {\em Physica}, A392:6341--6352, 2013.

\bibitem{PhysRevE.86.021130}
George Ruppeiner.
\newblock Thermodynamic curvature from the critical point to the triple point.
\newblock {\em Phys. Rev. E}, 86:021130, Aug 2012.

\bibitem{PhysRevE.85.031201}
Helge-Otmar May and Peter Mausbach.
\newblock Riemannian geometry study of vapor-liquid phase equilibria and
  supercritical behavior of the lennard-jones fluid.
\newblock {\em Phys. Rev. E}, 85:031201, Mar 2012.

\bibitem{PhysRevE.88.032123}
Helge-Otmar May, Peter Mausbach, and George Ruppeiner.
\newblock Thermodynamic curvature for attractive and repulsive intermolecular
  forces.
\newblock {\em Phys. Rev. E}, 88:032123, Sep 2013.

\bibitem{Dutta:2013dca}
Suvankar Dutta, Akash Jain, and Rahul Soni.
\newblock {Dyonic Black Hole and Holography}.
\newblock {\em JHEP}, 1312:060, 2013.

\bibitem{Lu:2013ura}
H.~Lü, Yi~Pang, and C.N. Pope.
\newblock {AdS Dyonic Black Hole and its Thermodynamics}.
\newblock {\em JHEP}, 1311:033, 2013.

\bibitem{Henneaux:1984ji}
M.~Henneaux and C.~Teitelboim.
\newblock {The cosmological constant as a canonical variable}.
\newblock {\em Phys.Lett.}, B143:415--420, 1984.

\bibitem{Banerjee:2011raa}
Rabin Banerjee, Sujoy~Kumar Modak, and Dibakar Roychowdhury.
\newblock {A unified picture of phase transition: from liquid-vapour systems to
  AdS black holes}.
\newblock {\em JHEP}, 1210:125, 2012.

\bibitem{SahayThesis:2010}
Anurag Sahay.
\newblock {\em {Thermodynamic geometry and critical phenomena of Black holes}}.
\newblock PhD thesis, Indian Institute of Technology Kanpur, Kanpur-208016,
  India, 2010.

\end{thebibliography}
\bibliographystyle{unsrt}

\end{document}